\documentclass[table]{SciPost}

\hypersetup{
    colorlinks,
    linkcolor={red!50!black},
    citecolor={blue!50!black},
    urlcolor={blue!80!black}
}

\usepackage[bitstream-charter]{mathdesign}
\DeclareSymbolFont{usualmathcal}{OMS}{cmsy}{m}{n}
\DeclareSymbolFontAlphabet{\mathcal}{usualmathcal}

\fancypagestyle{SPstyle}{
    \fancyhf{}
    \lhead{\colorbox{scipostblue}{\bf \color{white} ~SciPost Physics }}
    \rhead{{\bf \color{scipostdeepblue} ~Submission }}
    
    \fancyfoot[C]{\textbf{\thepage}}
}

\usepackage{xcolor}
\usepackage[disable]{todonotes}
\usepackage{siunitx}  
    \DeclareSIUnit{\angstrom}{\mbox{\normalfont{\AA}}}
\usepackage[capitalize]{cleveref}
\usepackage{acronym}

\newcommand*{\crbr}{CrBr\textsubscript{3}}
\newcommand*{\crcl}{CrCl\textsubscript{3}}
\newcommand*{\nbse}{NbSe\textsubscript{2}}
\newcommand*{\moire}{moiré}
\newcommand*{\eysr}{\ensuremath{E_{\text{YSR}}}}
\newcommand*{\qeff}{\ensuremath{Q_{\text{eff}}}}
\newcommand*{\charge}[1]{\ensuremath{#1\text{e}}}
\newcommand*{\didv}{dI/dV}
\newcommand*{\Vb}{\ensuremath{V_{\text{b}}}}
\newcommand*{\Vac}{\ensuremath{V_{\text{ac}}}}

\newacro{stm}[STM]{scanning tunneling microscopy}
\newacro{sts}[STS]{scanning tunneling spectroscopy}
\newacro{dos}[DOS]{density of states}
\newacro{cdw}[CDW]{charge density wave}
\newacro{ysr}[YSR]{Yu-Shiba-Rusinov}
\newacro{Ef}[$E_{\text{F}}$]{Fermi level}
\newacro{mzm}[MZM]{Majorana zero-mode}
\newacro{tsc}[TSC]{topological superconductivity}
\newacro{zbp}[ZBP]{zero-bias peak}
\newacro{mbe}[MBE]{molecular beam epitaxy}
\newacro{hemt}[HEMT]{high-electron-mobility transistor}

\begin{document}

\pagestyle{SPstyle}

\begin{center}{\Large \textbf{\color{scipostdeepblue}{
    Non-Topological Edge-Localized \acl{ysr} States in \crbr{}/\nbse{} Heterostructures
}}}\end{center}

\begin{center}
    \textbf{
        Jan P. Cuperus,
        Daniel Vanmaekelbergh and
        Ingmar Swart\textsuperscript{$\star$}
    }
\end{center}

\begin{center}
    Debye Institute for Nanomaterials Science,
    Utrecht University,
    The Netherlands
    \\[\baselineskip]
    $\star$ \href{mailto:i.swart@uu.nl}{\small i.swart@uu.nl}
\end{center}

\section*{\color{scipostdeepblue}{Abstract}}
\textbf{%
    \boldmath{%
    Topological superconductivity is predicted to emerge in certain magnet-superconductor hybrid systems.
    Here, we revisit the heterostructure of insulating monolayer \crbr{} and \nbse{}, for which different conclusions on the presence of topological superconductivity have been reported.
    Using low-temperature scanning tunneling microscopy and (shot noise) spectroscopy, we find that the superconducting gap well inside the \crbr{} islands is not affected by magnetism.
    At the island edges, we observe \ac{ysr} states at a variety of in-gap positions, including zero energy.
    The absence of topological superconductivity is verified by extensive \didv{} measurements at the \crbr{} island edges.
    Our results ask for a more detailed understanding of the interaction between magnetic insulators and superconductors.
    }
}

\vspace{\baselineskip}

\noindent\textcolor{white!90!black}{%
\fbox{\parbox{0.975\linewidth}{%
\textcolor{white!40!black}{\begin{tabular}{lr}%
  \begin{minipage}{0.6\textwidth}%
    {\small Copyright attribution to authors. \newline
    This work is a submission to SciPost Physics. \newline
    License information to appear upon publication. \newline
    Publication information to appear upon publication.}
  \end{minipage} & \begin{minipage}{0.4\textwidth}
    {\small Received Date \newline Accepted Date \newline Published Date}%
  \end{minipage}
\end{tabular}}
}}
}


\vspace{10pt}
\noindent\rule{\textwidth}{1pt}
\tableofcontents
\noindent\rule{\textwidth}{1pt}
\vspace{10pt}

\section{\label{sec:intro}Introduction}
At present, the road to scalable quantum computing is cumbersome due to the qubit decoherence problem~\cite{shorScheme1995,chaeelementary2024}.
Despite recent advancements in quantum error correction, the decoherence problem requires large physical and technological overhead~\cite{acharyaSuppressing2023,acharyaQuantum2024}.
The development of topological qubits, which have significantly lower decoherence rates, would therefore be a leap forward towards scalable quantum computing.
Central to the working principle of topological qubits are \acp{mzm}, exotic fermionic particles that obey non-Abelian statistics~\cite{aliceaNew2012,beenakkerSearch2013}.
As a potential platform for the realization of \acp{mzm}, topological superconductors are actively sought after in condensed matter systems.
Examples of material systems that have shown signs of \acp{mzm} include proximitized semiconductor nanowires~\cite{oregHelical2010,lutchynMajorana2010,mourikSignatures2012,pradaAndreev2020}, proximitized topological insulators~\cite{fuSuperconducting2008,wangcoexistence2012,sunMajorana2016,flotottoSuperconducting2018}, superconducting topological surface states~\cite{wangEvidence2018,machidaZeroenergy2019}, and 1D/2D magnetic lattices on superconducting surfaces~\cite{pientkaTopological2013,nadj-pergeProposal2013,liTwodimensional2016,nadj-pergeObservation2014,menardTwodimensional2017,palacio-moralesAtomicscale2019,schneiderTopological2021}.
In the examples above, \ac{tsc} is established by the induced, effective p-wave pairing of superconductivity; the required spin degeneracy lifting is provided by a topological phase transition~\cite{aliceaNew2012}, as has been shown for lattices of \ac{ysr} states~\cite{schneiderTopological2021}.
\ac{ysr} states are in-gap bound states induced by the exchange interaction between magnetic atoms (historically: impurities) and Cooper pairs~\cite{yuBound1965,shibaClassical1968,rusinovSuperconductivity1969,rusinovTheory1969}.
In the classical spin approximation~\cite{heinrichSingle2018}, the energy of \ac{ysr} states (\eysr{}) with respect to the \ac{Ef} depends on the exchange coupling $J$ and the impurity spin $S_{\text{imp}}$:

\begin{equation}
    \label{eq:ysr-energy}
    \eysr{} = \Delta \frac{1 - a^2}{1 + a^2}
    \quad \text{with} \quad
    a = J S_{\text{imp}} \pi \rho_s.
\end{equation}
Here, $\Delta$ is the superconducting gap size and $a$ parameterizes the impurity-substrate interaction via $J$, $S_{\text{imp}}$, and $\rho_s$, which is the density of states of the substrate at \ac{Ef} in the normal state.
From \cref{eq:ysr-energy}, it follows that \ac{ysr} states may be localized at \ac{Ef}, in which case they would be indistinguishable from a \ac{mzm} in density of states measurements.
This implies that a \ac{zbp} in a density of states measurement is no definitive proof of \acp{mzm} (or \ac{tsc}).
Other arguments are thus required to label a \ac{zbp} as a \ac{mzm}.
For example, plateaus at quantized conductance values in transport experiments would support a \ac{mzm} claim~\cite{lawMajorana2009,wimmerQuantum2011}.

Recently, the existence of a \ac{zbp} at the edges of \crbr{} islands, grown on \nbse{}, has been interpreted as a 1D \ac{mzm}~\cite{kezilebiekeTopological2020}.
It is reasoned that the magnetic insulator \crbr{} interacts with the $s$-wave superconductivity of the \nbse{} substrate, leading to \ac{ysr} bands that are in a topological phase, thanks to a moiré potential~\cite{kezilebiekeMoireEnabled2022}.
The \ac{zbp} observation is supported by an extensive theoretical model, which can also explain the observed interruptions of the 1D \ac{mzm}.
Given the impact of this finding, we set out to confirm these observations and improve the understanding of this exciting material platform.
Here, we report independent \acs{stm} measurements of \crbr{}/\nbse{} heterostructures, grown by \acs{mbe}.
In contrast to refs.~\cite{kezilebiekeTopological2020,kezilebiekeMoireEnabled2022}, we find no \ac{ysr} bands in the interior of the \crbr{} islands, and no 1D \ac{zbp} at the edges.
Instead, we find edge-localized \ac{ysr} states, whose energies are spread across the superconducting gap (including very close to zero energy).
The \ac{ysr} states disperse along small edge segments, and the \ac{ysr} state energy can be tuned by interaction with the \acs{stm} tip.
Our findings are in excellent agreement with the recent report of Li et al.~\cite{liObservation2024}, which was published shortly after our experiments had been performed.

\section{\label{sec:methods}Methods}
\subsection*{Sample preparation}
The growth of \crbr{} thin films was performed in an ultra-high vacuum chamber with a base pressure of ca. \qty{5e-10}{mbar}.
A \nbse{} single crystal (HQGraphene), glued to a molybdenum sample holder with conductive epoxy (EpoTek H20E), was used as the substrate.
The \nbse{} surface was prepared by cleaving, using scotch tape, in the fast entry lock (pressure lower than \qty{4e-8}{mbar}).
Prior to cleaving, the \nbse{} sample was degassed up to \qty{300}{\degreeCelsius}.
\crbr{} was evaporated from an e-beam evaporator (EFM3, Focus GmbH) containing a \crbr{} single crystal (HQGraphene) in a quartz cup.
A sub-monolayer coverage of crystalline \crbr{} islands was obtained by evaporating for \qty{6}{min} onto a freshly cleaved \nbse{} crystal, which was heated to \qty{230}{\degreeCelsius}.
After growth, the sample was kept at the growth temperature for \qty{15}{min}.
After cooldown, the sample was transferred to the STM head, which is part of the same ultra-high vacuum system.

\subsection*{STM/STS measurements}
All \ac{stm} and \ac{sts} measurements were performed using a USM1300 (Unisoku Co. Ltd.), operated at $T = \qty{360}{mK}$.
A mechanically polished PtIr wire, conditioned on Au(111), was used as the STM tip.
In \ac{sts} experiments, \didv{} signals were recorded by lock-in detection, using a modulation frequency of either \qty{771}{Hz} or \qty{971}{Hz}.
Modulation amplitudes \Vac{} are given in the description of each experiment.
STM and STS data was organized using SPMImageTycoon~\cite{rissSpmImage2022}; STM images were analyzed using Gwyddion~\cite{Necas2012}.

\subsection*{Shot noise measurements}
We have measured the shot noise generated by the tunneling current at \unit{MHz} frequency using a recently developed amplification circuit~\cite{bastiaansAmplifier2018}.
In essence, the circuit consists of a bias tee, a superconducting LC resonator ($f_0 = \qty{2.8}{MHz}$, $Q \approx \num{1800}$), and a \ac{hemt}.
The bias tee is used to separate low frequency signals (STM feedback, \didv{} signal) from high frequency signals, whereas the LC resonator converts current noise from the junction into voltage noise at the gate of the HEMT.
Lastly, the \ac{hemt} is used to match the high impedance circuit to the \qty{50}{\ohm} impedance of a transmission line.
All these components are located in the vicinity of the \qty{1}{K} pot, such that their temperature is expected to be $< \qty{2}{K}$ during the measurements at \qty{360}{mK} ($T_{\text{1K}} \sim \qty{1.6}{K}$).
At room temperature, the noise signal is amplified by a \qty{+40}{dB} current amplifier (Femto HSA-X-1-40), before it is measured by a digital spectrum analyzer (MFLI, Zurich Instruments AG).

In a shot noise spectroscopy experiment, the magnitude of shot noise is measured at a set of bias voltages.
In this report, shot noise spectroscopy was performed with the STM feedback configured to maintain a constant junction resistance $R_{\text{J}} = \Vb{} / I$.
We note that this implies a non-constant transparency of the tunnel junction, in the case of a non-constant density of states (such as a SC gap).
To find the shot noise magnitude $S_{\text{I}}$, the power spectral density $S_{\text{V}}(\omega)$ is measured in a \qty{100}{kHz} bandwidth around the resonator's center frequency, using lock-in detection (zoom FFT).
For each bias point, the current noise is extracted from a fit of $S_{\text{V}}(\omega)$ using a simplified model of the circuit:
\begin{equation}
    \label{eq:sn-psd}
    S_\text{V}(\omega) = S_{\text{tot}} \times |Z_{\text{tot}}(\omega)|^2 + S_{\text{V}}^0\text{,}
\end{equation}
where $Z_{\text{tot}}(\omega)$ is the complex impedance of the resonator circuit, in parallel with the junction resistance $R_\text{J}$, and $S_{\text{V}}^0$ is the background (input) voltage noise level.
The resonator circuit is approximated by two parallel, identical RLC resonators ($R \sim \qty{1}{\ohm}, L = \qty{120}{\micro \henry}, C \sim \qty{26}{pF}$), separated by a coupling capacitor ($C_\text{c} = \qty{100}{pF}$).
The fit parameters are $S_\text{tot}$, $S_\text{V}^0$, $R$ and $C$.
$S_{\text{tot}} = G^2 \left(S_{\text{I}} + S_{\text{I}}^0\right)$ is the total current noise, with $G$ being the gain of the amplification chain and $S_{\text{I}}^0$ being the background current noise level.
The effective charge $Q_{eff}$ is extracted from the obtained $S_{\text{tot}} (\Vb{})$ values using
\begin{equation}
    \label{eq:sn-qeff}
    S_{\text{tot}} = G^2 \left[ 2 \qeff |I| \times \mathrm{coth}\left(\frac{\qeff{}\Vb{}}{2k_{\text{B}}T}\right) + S_{\text{I}}^0 \right]\text{,}
\end{equation}
where the Schottky formula for shot noise is recognized as $2 \qeff |I|$, and the $\mathrm{coth}$-term accounts for thermal equilibrium noise of the tunnel junction~\cite{birkShotNoise1995}.
\Cref{eq:sn-qeff} is fitted to $S_{\text{tot}} (\Vb{})$ for bias points where the effective charge \qeff{} is known to be \charge{1}, i.e., outside the SC gap.
We find a gain $G$ of \num{0.176} (excluding room temperature amplification), whereas $S_{\text{I}}^0$ is ca. \qty{50}{fA^2/Hz}.
In our analysis, the thermal noise $S_{\text{I}}^{\text{th}} = 4k_{\text{B}}T/Z_{\text{tot}}$ is included in $S_{\text{I}}^0$ and not taken into account in the resonance circuit fit.
By measuring each $S_{\text{V}}(\omega)$ for ca. \qty{10}{min} (\num{4000} FFT repetitions), we can determine $S_I$ to within \qty{0.15}{fA^2/Hz} (\qty{95}{\percent} confidence interval), independent of current/bias.
This uncertainty is taken into account in the calculation of \qeff{}, for which the confidence interval becomes larger for smaller currents, because of a lower signal-to-noise ratio.

\section{\label{sec:results}Results and Discussion}
\subsection{Morphology \& Structure}
\begin{figure}[h]
    \includegraphics[width=\textwidth]{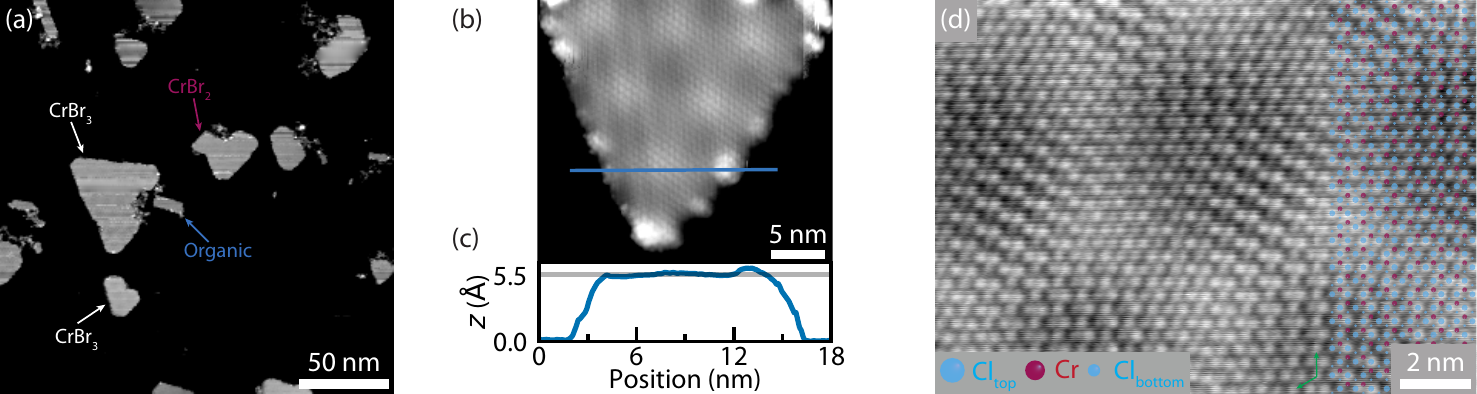}
    \caption{
        \label{fig:structure}
        Structure of as-grown \crbr{}/\nbse{} heterostructures at different length scales.
        (a) Morphology of the sample, showing that the \crbr{} islands of different sizes coexist with small patches of CrBr\textsubscript{2} and some organic residue.
        Set point: \qty{+0.9}{V}, \qty{50}{pA}.
        (b) Image of a single \crbr{} island, on which both the atomic and moiré unit cells can be recognized.
        Set point: \qty{+1.5}{V}, \qty{50}{pA}.
        (c) Line profile measured along the blue line shown in (b).
        The apparent height of the \crbr{} island is \qty{5.5}{\angstrom}.
        (d) Image of the \crbr{} lattice with atomic resolution: individual Br atoms are visible, as well as the moiré pattern.
        An atomic model of the \crbr{} lattice is overlaid.
        Set point: \qty{+0.93}{V}, \qty{50}{pA}.
        }
\end{figure}

\Cref{fig:structure}a shows the typical morphology of the as-grown \crbr{}/\nbse{} heterostructure.
\crbr{} has formed monolayer islands of triangular shape, which vary in size from \qty{10}{nm} to \qty{60}{nm}.
The islands have an apparent height of $\sim \qty{5.5}{\angstrom}$ at $\Vb{} = \qty{+1.5}{\volt}$ (\qty{5.0}{\angstrom} at $\Vb{} = \qty{0.9}{\volt}$), indicative of their monolayer thickness.
In addition to the \crbr{} islands, the evaporation process has introduced some organic contaminants, visible as elongated structures (annotated in blue).
Small patches of CrBr\textsubscript{2} are also observed (annotated in red), always attached to \crbr{} islands and indicative of slight \crbr{} degradation~\cite{kezilebiekeElectronic2021}.

On the \crbr{} islands, a \moire{} pattern is visible, resulting from the lattice mismatch between \crbr{} and \nbse{} (see \cref{fig:structure}b).
The \moire{} pattern has a periodicity of $\qty{6.1}{nm} \pm \qty{0.1}{nm}$, which is similar to what has been reported previously~\cite{kezilebiekeTopological2020,liObservation2024}.
Furthermore, \cref{fig:structure}b shows that the \crbr{} island is crystalline in the interior, but that the edges are not atomically straight -- also in line with the previous reports.
In an \ac{stm} topograph with atomic contrast, shown in \cref{fig:structure}c, the \moire{} lattice can be observed together with the characteristic atomic pattern of \crbr{}.
The moiré pattern is not exactly aligned with the rows of Br atoms, showing that the twist angle is not exactly \qty{30}{\degree} (as suggested in previous reports), but more close to \qty{32}{\degree}.
Additionally, we observed that \crbr{} islands on adjacent \nbse{} planes, which have mirrored sublattice symmetries, have a mirrored orientation (see \cref{fig:mirrored-orientation}).
This indicates that the twist angle is (in part) determined by the atomic interaction between the substrate and the epilayer.

\subsection{Electronic Characterization}
We have investigated how the magnetic insulator \crbr{} affects the electronic structure of \nbse{} via \ac{sts} experiments inside the band gap of \crbr{}.
By measuring inside the band gap, \ac{sts} experiments can probe the \nbse{} \ac{dos} beneath the \crbr{} islands.
To study charge transfer effects, we recorded \ac{sts} data in the energy range from $\Vb{} = \qty{-0.5}{V}$ to $\Vb{} = \qty{+0.75}{V}$.
By comparing data acquired on the \crbr{} islands to data recorded on \nbse{}, we have found that the \crbr{} induces a shift of ca. \qty{+30}{mV} of the Nb d-band at $\Vb{} \sim \qty{+300}{mV}$.
This shift has been interpreted as a sign of electron doping from \nbse{} to \crbr{} and is, given the uncertainty, similar in magnitude to what has been reported by refs.~\cite{kezilebiekeTopological2020} and \cite{liObservation2024}\footnotemark{}.
\footnotetext{Ref.~\cite{kezilebiekeTopological2020} reports a shift of $+\qty{80}{mV}$ based on a contour plot, but from the line plot a value of $+\qty{30}{mV}$ seems more appropriate.}

\begin{figure}
    \centering
    \includegraphics[width=\textwidth]{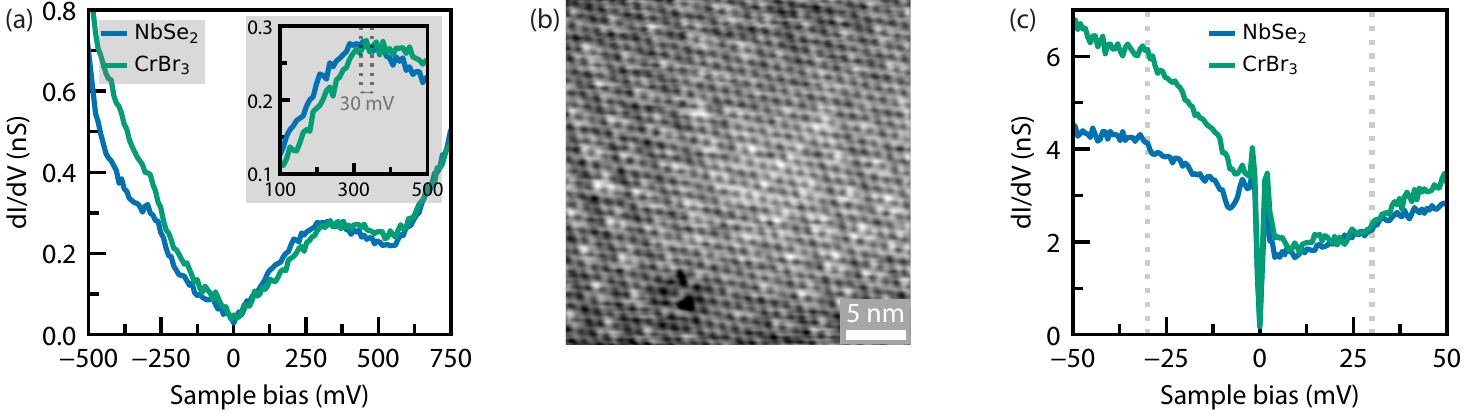}
    \caption{
        \label{fig:large-sts}
        Characterization of the electronic structure of \crbr{}/\nbse{}.
        (a) \didv{} spectrum probing the charge transfer from \nbse{} to \crbr{}.
        The Nb d-band at \qty{+300}{mV} shifts by ca. \qty{+30}{mV}, indicative of slight electron doping of \crbr{}.
        Set point: \qty{1}{V}, \qty{200}{pA}, $\Vac{} = \qty{9}{mV}$.
        (b) \ac{stm} image showing the \ac{cdw} of \nbse{}.
        Set point: \qty{+200}{mV}, \qty{100}{pA}.
        (c) \didv{} spectrum probing the \ac{cdw} of \nbse{} beneath a \crbr{} island.
        Set point: \qty{+100}{mV}, \qty{300}{pA}, $\Vac{} = \qty{0.65}{mV}$.
        }
\end{figure}

In the smaller bias range from $\Vb{} = \qty{-100}{mV}$ to $\Vb{} = \qty{+100}{mV}$, we have used \ac{sts} experiments to probe the \ac{cdw} beneath the \crbr{} islands.
Away from the islands, the \ac{cdw} is readily visible in \ac{stm} topographs such as \cref{fig:large-sts}{b}.
In \ac{sts}, the \ac{cdw} is manifested as kinks at ca. $\pm \qty{35}{mV}$~\cite{soumyanarayananQuantum2013}.
As shown in \cref{fig:large-sts}c, we indeed observe a change of slope around $\pm \qty{30}{mV}$, both on \nbse{} (blue curve) and on \crbr{} (green curve).

\subsection{SC Gap Characterization}
\begin{figure}
    \centering
    \includegraphics[width=\textwidth]{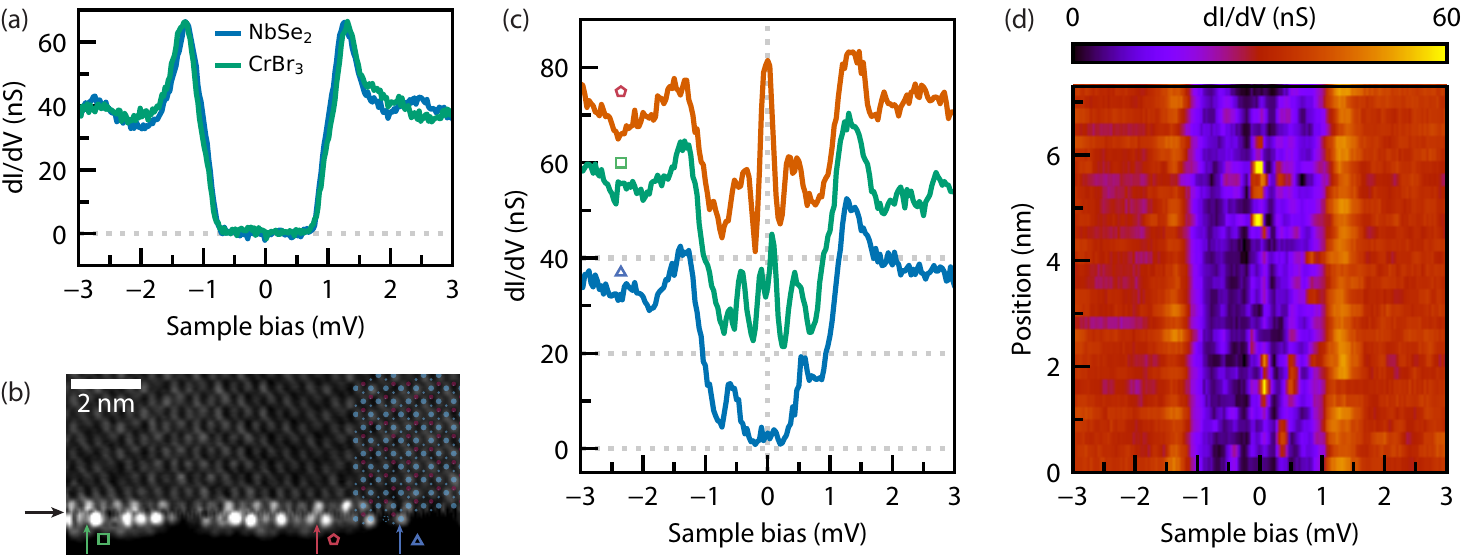}
    \caption{
        \label{fig:sc-gap-sts}
        (a) SC gap of \nbse{} (green) and \crbr{} (blue).
        A hard SC gap is observed on both locations.
        (b) Constant height image of the edge segment probed by the STS experiments shown in (c), (d).
        Set point: \qty{+5}{mV}, \qty{50}{pA}.
        (c) Selected spectra from the line spectrum shown in (d), displaying the variety of in-gap profile found on a single edge segment.
        Locations of the spectra are indicated in (b).
        Spectra are offset vertically by \qty{20}{nS} for clarity.
        (d) Contour plot of \didv{} spectra acquired along a line on the edge shown in (b).
        The location of the line spectrum is indicated by the black horizontal arrow in (b).
        Set points for (a), (c) and (d): \qty{+5}{mV}, \qty{150}{pA}, $\Vac{} = \qty{33}{\micro V}$.
        }
\end{figure}

To investigate the effect of \crbr{} on the superconductivity of \nbse{}, we performed \ac{sts} experiments in a small bias window around \ac{Ef}.
First, we compared the SC gap measured on the \crbr{} islands to measurements on bare \nbse{}.
We find that the SC gaps measured on both locations are nearly identical, as shown in \cref{fig:sc-gap-sts}a.
In contrast to ref.~\cite{kezilebiekeTopological2020}, but in line with ref.~\cite{liObservation2024}, no \ac{ysr} bands are found in the interior of the \crbr{} islands.
As a more detailed analysis, we have fitted the data of several \nbse{} and \crbr{} SC gaps using the McMillan model~\cite{mcmillanTunneling1968,noatQuasiparticle2015}, but we do not observe a trend in the fit parameters indicative of \ac{ysr} bands (see \cref{sec:sc-gap-fits}).
An investigation of the superconductivity in the interior of the \crbr{} islands by shot noise spectroscopy is presented below.

Next, we focused on the edges of the \crbr{} islands.
By recording constant height maps, we have mapped the atomic structure of an edge (see \cref{fig:sc-gap-sts}b).
In the interior of the island, the Br sublattice can be observed, from which we can extract the orientation of the \crbr{} lattice (see the atomic model overlay in \cref{fig:sc-gap-sts}b).
From the orientation, it follows that the edge corresponds to the armchair edge of the honeycomb sublattice formed by the Cr atoms.
Even in close vicinity of the edge, the \crbr{} crystallinity is pristine; only in the last unit cell ($\sim \qty{6}{\angstrom}$), the atomic structure is more disordered, and in some parts, missing atoms can be identified.
However, in other edge segments, the \crbr{} lattice appears intact, except for the unavoidable termination of the periodicity.
In such parts, the outermost Br atoms are only faintly visible.
The low intensity of these Br atoms suggests that they are bent downwards (towards the substrate), although we cannot rule out that this effect stems from the electronic structure instead.
Either way, the presence of these Br atoms implies that the Cr atoms at the edge are sixfold coordinated, as they would be in the interior of the \crbr{} island.

The constant-height current map (\cref{fig:sc-gap-sts}b) shows clear intensity variations along the edge.
To explain this observation, we turn to the spectroscopy data acquired on this edge.
\didv{} spectra acquired on positions that have different intensities in \cref{fig:sc-gap-sts}b exhibit very different in-gap features.
The in-gap features come in pairs, which appear symmetrically around \ac{Ef}, but are asymmetric in intensity.
Based on the energy symmetry, we interpret the peak pairs to be \ac{ysr} states.
We note that these \ac{ysr} states most likely do not result from undercoordinated Cr atoms at the edge, as the atomic structure is pristine (as discussed above).
Instead, we expect the \ac{ysr} states to arise from a Cr $3d$ \ac{dos} at \ac{Ef} that is increased at the edge, as we have reported for \crcl{}/\nbse{}~\cite{cuperusOne2025}.
The increased \ac{dos} strengthens the exchange interaction between the Cr atoms and the \nbse{} substrate, and can explain the edge-localization of the \ac{ysr} states.

Depending on position, the main \ac{ysr} states can be located close to the coherence peaks (bottom curve), close to \ac{Ef} (middle curve), or precisely at \ac{Ef} (upper curve).
In the latter case, the \ac{ysr} states are so close in energy that only a single peak is observed at \ac{Ef}.
This happens despite the fact that in-gap states observed here have a smaller full-width at half maximum than those reported in ref.~\cite{kezilebiekeTopological2020}.
Not only the energy, but also the intensity of the \ac{ysr} states can vary greatly, independent on the \ac{ysr} state energy.
To characterize the edge-localized \ac{ysr} states more systematically, we have recorded a line spectrum along the edge (see the contour plot in \cref{fig:sc-gap-sts}d).
The line spectrum shows that the \ac{ysr} state energies vary smoothly with position, indicative of coupling between the individual \ac{ysr} states.
The coupling also extends across the low-intensity segment at ca. \qty{3}{nm} from the left.
The fact that the peak positions vary spatially, and are not pinned to zero energy, rules out the presence of topological superconductivity in our experiments.
Instead, the \ac{ysr} states are more likely to originate from the (magnetic) Cr atoms at the edge, similar to what we have observed for \crcl{}/\nbse{}~\cite{cuperusOne2025}.

\begin{figure}
    \centering
    \includegraphics{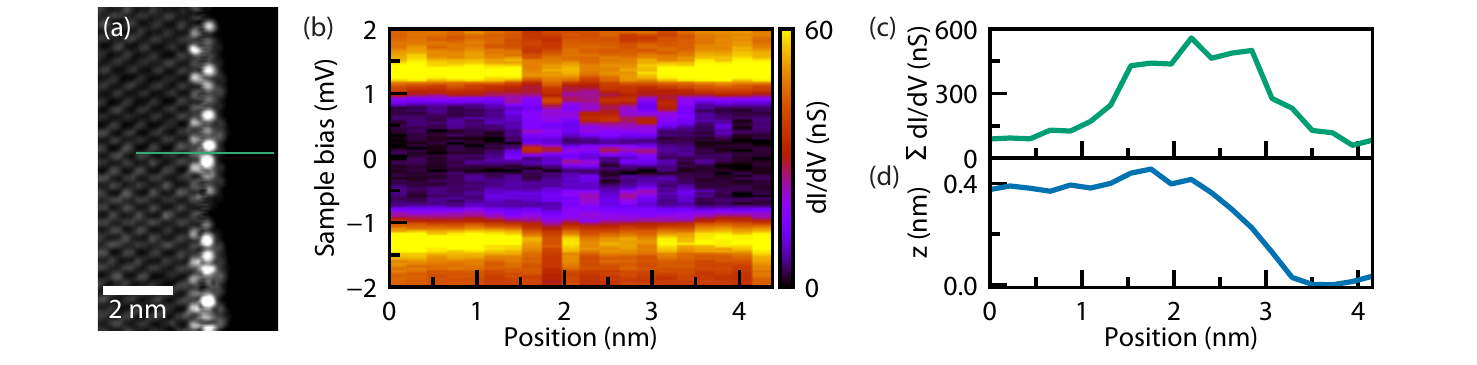}
    \caption{
        \label{fig:ls-perpendicular}
        (a) Constant height map of a \crbr{} island edge.
        The green line indicates the position of the line spectrum, which is shown in (b).
        Set point: \qty{+5}{mV}, \qty{50}{pA}.
        (b) Line spectrum measured across a \crbr{} edge.
        In-gap states are observed in a \qty{2}{nm} wide region close to the edge.
        Set point: \qty{+5}{mV}, \qty{150}{pA}, $\Vac{} = \qty{33}{\micro \volt}$
        (c), (d) Summed in-gap \didv{} signal and tip height of the line spectrum shown in (b).
    }
\end{figure}

The constant height map of \cref{fig:sc-gap-sts}b suggests that the \ac{ysr} states are localized to region of width $\sim \qty{1}{nm}$ from the edge.
To confirm this edge-localization, we performed \didv{} spectroscopy on a line perpendicular to the edge.
Also in this line spectrum, the in-gap states appear within a small region close to the edge (see \cref{fig:ls-perpendicular}b).
Compared to the high-intensity region in the constant height map of \cref{fig:ls-perpendicular}a, the \ac{ysr} states in the line spectrum appear slightly more extended to the outside of the island.
We attribute this to the fact that the feedback is enabled between each position in the line spectrum.

\subsection{Further Characterization of YSR edge states}\label{sec:further-char-ysr}
\begin{figure}
    \centering
    \includegraphics[width=\textwidth]{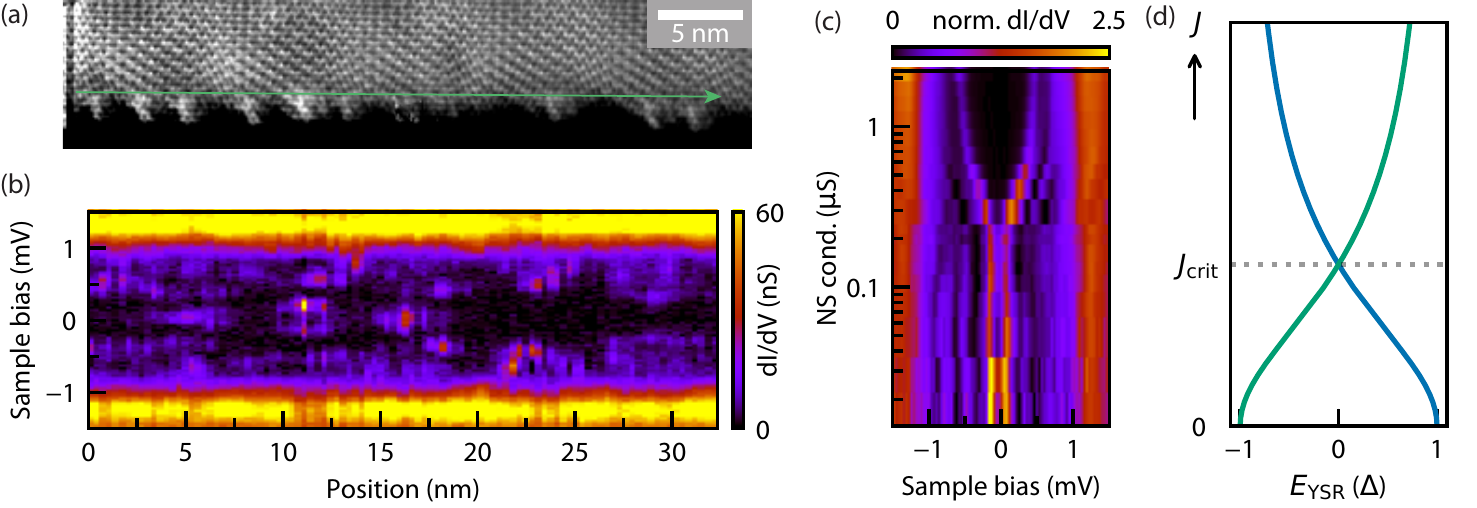}
    \caption{
        \label{fig:more-char}
        (a) Constant height \ac{stm} image of an island edge that is longer but more disordered than the edge shown in \cref{fig:sc-gap-sts}b.
        Set point: \qty{+5}{mV}, \qty{50}{pA}.
        (b) Contour plot of \didv{} spectra acquired along the green line shown in (a).
        In-gap states disperse along small edge segments only.
        Set point: \qty{+5}{mV}, \qty{150}{pA}, $\Vac{} = \qty{40}{\micro V}$.
        (c) Contour plot of \didv{} spectra of a \ac{ysr} state, acquired at different tip-sample distances.
        By approaching the sample, the \ac{ysr} state energy is tuned through the quantum phase transition.
        The normal-state (NS) conductance is defined as the average conductance signal in the bias range $|V_b| = \qtyrange{3}{5}{mV}$.
        To calculate the normalized conductance, the differential conductance is divided by the NS conductance.
        Set points: \qty{+5}{mV}, 70 pA to 9 nA, $\Vac{} = \qty{30}{\micro V}$.
        (d) Position of the \ac{ysr} state energy as a function of the exchange interaction $J$, according to \cref{eq:ysr-energy}.
        The quantum phase transition occurs at $J = J_{\text{crit}}$.
        }
\end{figure}

\Ac{ysr} states are sensitive to the chemical environment of the magnetic species.
For example, it has been shown that the energy of YSR states is affected by the \ac{cdw} of \nbse{}~\cite{liebhaberYu2020} and by the moiré pattern of self-assembled molecular monolayers~\cite{hatterScaling2017}.
Here, we observed the \ac{ysr} states at the \crbr{} edges to be highly sensitive to the surrounding environment.
In the \crbr{} edge (line spectrum) shown in \cref{fig:sc-gap-sts}, the edge is well-ordered, resulting in an in-gap profile that changes smoothly along the edge.
However, at an edge that is only slightly more disordered, we observed a variety of in-gap profiles, as shown in \cref{fig:more-char}a, b.
Spatially dispersive \ac{ysr} states can still be recognized along small edge segments, but completely different in-gap profiles appear in adjacent segments.
Most notably, the in-gap profile contains \ac{ysr} states close to zero energy in the left part of the edge, whereas \ac{ysr} states are only found closer to the coherence peaks in the right part.

In addition to spatial variations, the \ac{ysr} state energy could also be altered by interaction with the \ac{stm} tip.
By recording the SC gap at varying tip-sample distances (i.e., different tunneling transmissivity), the \ac{ysr} state could be tracked through the quantum phase transition~\cite{heinrichSingle2018}, as shown in \cref{fig:more-char}c.
At the largest tip-sample distance (lowest transmissivity, bottom of \cref{fig:more-char}c), the most intense \ac{ysr} states are found at $\pm \qty{0.2}{meV}$.
Upon decreasing the tip-sample distance (increasing transmissivity), these states shift towards \ac{Ef}, up until a normal state conductance of $\sim \qty{0.3}{\micro S}$, after which the states split again.
By changing the tip-sample distance, we tune the atomic forces between tip, substrate and epilayer.
Presumably, the approaching tip pushes down the \crbr{}, thereby increasing the exchange coupling $J$ between the Cr atom(s) and the superconducting substrate~\cite{brandElectron2018,farinacciTuning2018,huangQuantum2020}.
The exchange coupling $J$ directly affects the \ac{ysr} state energy \eysr{} according to \cref{eq:ysr-energy}, which is visualized in \cref{fig:more-char}d.
In the weak coupling regime (small $J$, $a^2 < 1$), the ground state of the spin-superconductor system is a free-spin state.
In this regime, the in-gap states observed at positive (negative) energy in \ac{sts} experiments correspond to the excitation to the screened-spin state via the electron-like (hole-like) part of the \ac{ysr} state.
When the exchange coupling is increased beyond a certain magnitude $J_{\text{crit}}$, the system goes through a quantum phase transition, after which the screened-spin state becomes the ground state.
Beyond the phase transition, the electron- and hole-like components of the \ac{ysr} state have switched sides, i.e., the electron-like part is found at negative energy, and the hole-like part at positive energy.
In the \ac{sts} data of \cref{fig:more-char}c, this side-switching is visible by the reversal of the in-gap states' relative spectral weight.
At the largest tip-sample distance, the in-gap state at negative energy is most intense, while the most intense state is found at positive energy for the smallest tip-sample distances.
As may be clear from \cref{fig:more-char}d, \ac{ysr} states are located at \ac{Ef} when $J$ is tuned to be at the quantum phase transition ($J = J_\text{crit}$).
In the current experiment, however, the steps in tip-sample distance were too coarse to observe this point of the phase diagram.
Nevertheless, this experiment shows the tunability of the \ac{ysr} states found at the edges of \crbr{} islands.

\subsection{Further Characterization of SC gap on \crbr{} by Shot Noise Spectroscopy}
\begin{figure}
    \centering
    \includegraphics[width=\textwidth]{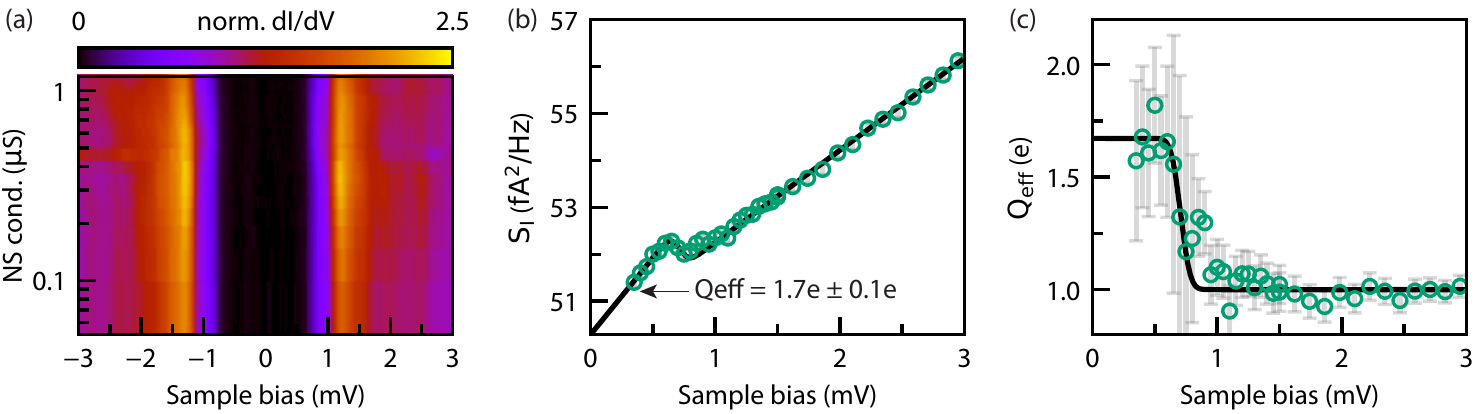}
    \caption{
        \label{fig:crbr-extra}
        (a) Contour plot of \didv{} spectra of a \ac{ysr} state, acquired at different tip-sample distances.
        The SC gap is not affected by the tip-sample distance.
        The NS conductance is calculated as in \cref{fig:more-char}b.
        Set points: \qty{+5}{mV}, \qty{300}{pA} - \qty{5}{nA}, $\Vac{} = \qty{30}{\micro V}$.
        (b), (c) Current noise $S_\text{I}$ (b) and effective charge \qeff{} (c) when tunneling into \crbr{}, as determined by shot noise spectroscopy.
        A constant junction resistance $R_\text{J} = \qty{5}{\mega \ohm}$ was maintained during the experiment.
        Inside the SC gap, an effective charge $\qeff{} = \charge{1.7} \pm \charge{0.1}$ is found.
        Error bars indicate \qty{95}{\percent} confidence intervals.
        }
\end{figure}

A main difference between our data and that of ref.~\cite{kezilebiekeTopological2020} is the absence of \ac{ysr} bands in the SC gap, as probed by \ac{sts} experiments inside the \crbr{} islands (see \cref{fig:sc-gap-sts}a).
To verify that the \ac{ysr} bands are truly absent (and not just very small), we characterized the SC gap on \crbr{} by means of regular \ac{sts} experiments at varying tip-sample distance (\cref{fig:crbr-extra}a) and by shot noise spectroscopy (\cref{fig:crbr-extra}b, c).
For the \ac{sts} experiments, the current set point (at \qty{5}{mV}) was increased from \qty{300}{pA} for the largest tip-sample distance to \qty{5}{nA} for the smallest tip-sample distance.
In the resulting \ac{sts} data, shown in \cref{fig:crbr-extra}a, no sign of \ac{ysr} bands is observed at any tip-sample distance.
Even at the smallest tip-sample distance, the SC gap is fully developed.
At negative bias, a small feature can be observed inside the SC gap.
We interpreted this feature as the wavefunction tail of one of the nearby ($\sim \qty{6}{nm}$) edge-localized \ac{ysr} states.
The extension of \ac{ysr} wavefunctions over such distances is possible in \nbse{}, thanks to the 2D character of its superconductivity~\cite{menardCoherent2015}.

In addition to these \ac{sts} experiments, we have performed shot noise spectroscopy to check for \ac{ysr} bands on \crbr{}.
In shot noise spectroscopy, the magnitude of shot noise, measured as a function of the bias voltage, is used to determine the effective charge of the tunneling (quasi)particles.
For a fully-developed superconducting gap, the effective charge changes from $\qeff{} = \charge{1}$ at bias voltages outside the gap to $\qeff{} = \charge{2}$ inside the gap, corresponding to a change in the tunneling mechanism from quasiparticle tunneling to Andreev reflections~\cite{dejongDoubled1994,jehlDetection2000,leflochDoubled2003,bastiaansImaging2019}.
In tunnel junction shot noise experiments, the effective charge $\qeff{}$ at in-gap biases is very sensitive to any \charge{1}-tunneling pathways available, such as quasiparticle poisoning by temperature or (residual) magnetic fields.
In constant conductance shot noise spectroscopy experiments, as performed here, a $\qeff{} = \charge{1}$ contribution of only \qty{1}{\percent} to the total tunnel current already reduces the effective charge from $\qeff{} = \charge{2}$ to $\qeff{} \approx \charge{1.1}$~\cite{geSingleelectron2023, geDirect2024}.
Here, we use this sensitivity to \charge{1}-tunneling pathways to search for \ac{ysr} bands that may be found in \crbr{}/\nbse{}.
In typical STM experiments, the excitation of a \ac{ysr} state is a single-electron process, which implies that tunneling events probing this excitation have an effective charge $\qeff{} = \charge{1}$~\cite{rubyTunneling2015}.
Hence, we expect to find an effective charge $\qeff{}$ well below \charge{2} on \crbr{}/\nbse{} if any \ac{ysr} bands are present~\cite{thupakulaCoherent2022}.
Instead, we find an effective charge of $\qeff{} = \charge{1.7} \pm \charge{0.1}$ (see \cref{fig:crbr-extra}b, c).
Following ref.~\cite{geSingleelectron2023}, this indicates that the contribution of \charge{1} tunneling events is smaller than \qty{0.1}{\percent}. 
The data in \cref{fig:crbr-extra} confirm what is suggested by \cref{fig:sc-gap-sts}: no \ac{ysr} bands exist in the \crbr{}/\nbse{} heterostructure.
Thus, the superconductivity is not affected by magnetic effects.

\section{\label{sec:conclusion}Conclusion}
Two recent scanning tunneling microscopy studies have reported contrasting findings on the existence of topological superconductivity in \crbr{}/\nbse{} heterostructures~\cite{kezilebiekeTopological2020,liObservation2024}.
Here, we have further characterized the \crbr{}/\nbse{} system by \ac{stm}, \ac{sts} and shot noise spectroscopy, and found no sign of topological superconductivity (in agreement with ref.~\cite{liObservation2024}).
The most notable difference with respect to the earlier work reporting topological superconductivity (ref.~\cite{kezilebiekeTopological2020}) is the observation of edge-localized \ac{ysr} states, instead of a peak pinned at zero energy.
Furthermore, in the interior of the \crbr{} islands, we observe a clean SC gap, essentially without quasiparticles ($< \qty{0.1}{\percent}$, as determined by shot noise spectroscopy).
The spatial profile of the \ac{ysr} states is reminiscent of the one we have recently observed in the related heterostructure \crcl{}/\nbse{}~\cite{cuperusOne2025}.
In analogy to the case of \crcl{}/\nbse{}, the edge-localized \ac{ysr} states presumably reflect an increase of the Cr-\nbse{} exchange interaction, which occurs even at perfect, unreconstructed \crbr{} edges.

At present, it is unclear what causes the absence (here, ref.~\cite{liObservation2024}) or presence (ref.\cite{kezilebiekeTopological2020}) of in-gap states in the interior of \crbr{} islands on \nbse{}.
The difference may be caused by a different substrate-epilayer interaction or, possibly, the presence of defects/impurities.
A systematic study into if/how the properties of \crbr{}/\nbse{} depend on the growth conditions may shed light on the origin of the different experimental observations.
Additionally, more insight into topological superconductivity in \crbr{}/\nbse{} may be obtained by the investigation of other magnetic 2D materials proximitized to $s$-wave superconductors.


\section*{Acknowledgments}
The authors thank P. Liljeroth and S. Kezilebieke for insightful discussions.

\paragraph{Author contributions}
I.S. conceived the project.
J.P.C. performed the experiments and analyzed the data.
D.V. and I.S. supervised the experiments.
All authors discussed the results and contributed to writing of the manuscript.

\paragraph{Conflicting interests}
The authors report no conflicting interests.

\paragraph{Funding information}
I.S. thankfully acknowledges funding by the European Research Council (Horizon 2020 “FRACTAL”, 865570).
D. V. and I. S. acknowledge the research program “Materials for the Quantum Age” (QuMat) for financial support.
This program (Registration Number 024.005.006) is part of the Gravitation program financed by the Dutch Ministry of Education, Culture and Science (OCW).

\paragraph{Data availability}
The data used to produce the figures in this manuscript have been published and can be downloaded from \cite{cuperusData2025}.
\clearpage

\renewcommand{\thefigure}{A\arabic{figure}}
\renewcommand{\thetable}{A\arabic{table}}
\setcounter{figure}{0}
\begin{appendix}
\section{Extended Data}

\subsection{Mirrored \crbr{} orientation on adjacent \nbse{} planes}
\begin{figure}[h]
    \centering
    \includegraphics{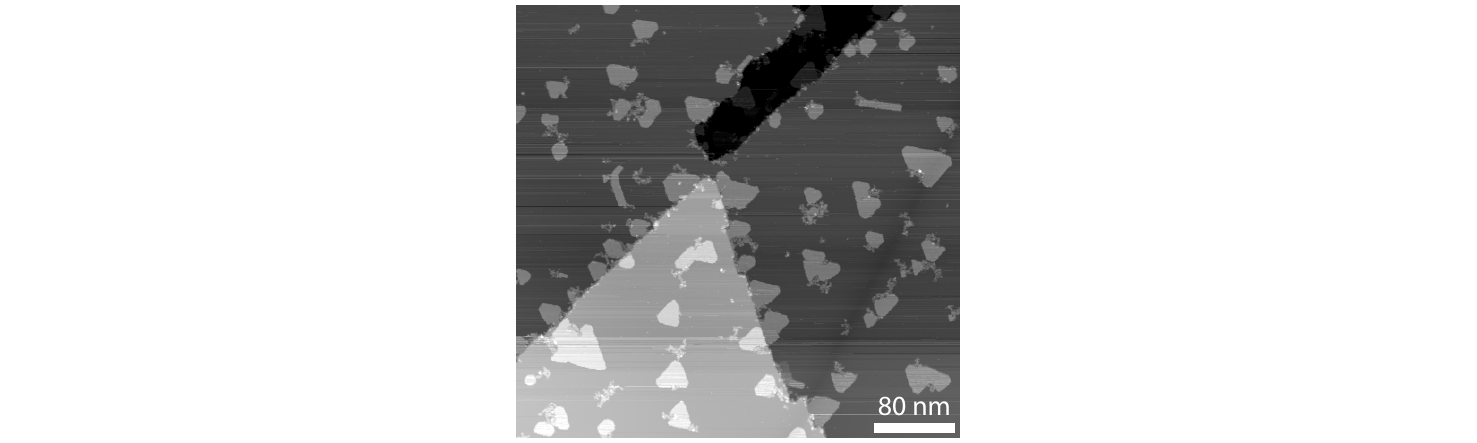}
    \caption{
        \label{fig:mirrored-orientation}
        \ac{stm} image of a location on the sample where different \nbse{} terraces are observed.
        \crbr{} islands on adjacent \nbse{} terraces have opposite orientations.
        Set point: \qty{0.9}{\volt}, \qty{50}{pA}.
    }
\end{figure}

\subsection{McMillan fitting of SC gap}\label{sec:sc-gap-fits}
\begin{figure}[p]
    \centering
    \includegraphics{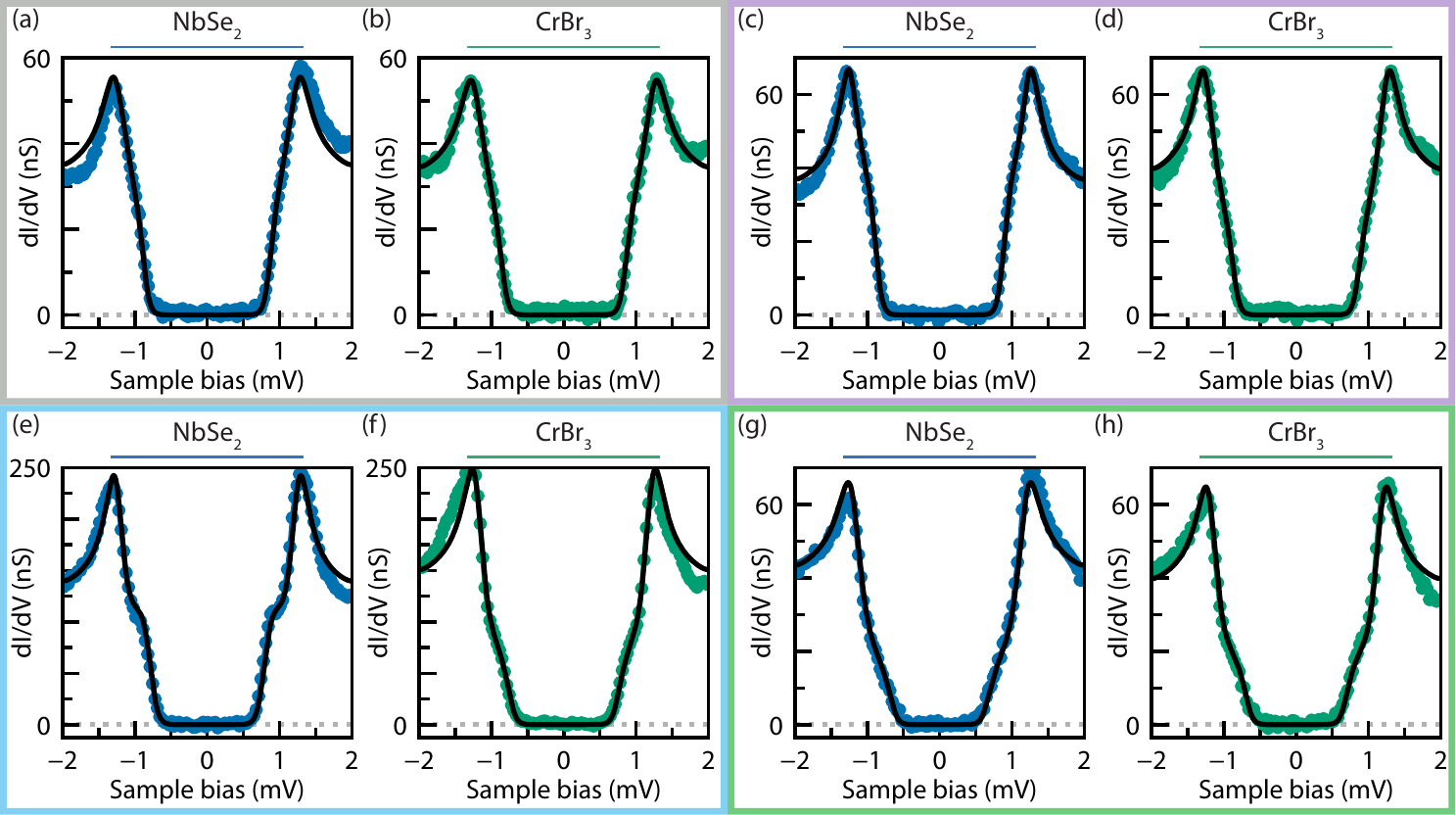}
    \caption{
        \label{fig:sc-gap-fits}
        \ac{sts} data of the SC gap on \nbse{} and \crbr{} positions, acquired on four different, macroscopic locations on the sample and with different nanoscopic tip.
        Fits to the McMillan model (\cref{eq:mcmillan,eq:gap-self-consistent}) are shown as black lines.
        Pairs of measurements on \nbse{} and \crbr{}, as indicated by the color coding, are performed consecutively, and with the same tip and the same measurement parameters.
        \ac{stm} feedback was disabled at \qty{5}{mV}, at a current set point of \qty{150}{pA} [(a)-(d), (g), (h)] or \qty{500}{pA} [(e), (f)].
        Lock-in amplitude $\Vac{}$ was either \qty{33}{\micro \volt} [(a)-(d)] or \qty{30}{\micro \volt} [(e)-(h)].
    }
\end{figure}

We have fitted the SC gap, as measured on different position on the sample and with different nanoscopic tips, using the McMillan model for two-band superconductivity.
Following ref.~\cite{noatQuasiparticle2015}, the \didv{} signal $G_\text{tot}$ was fitted to:
\begin{equation}\label{eq:mcmillan}
    G_\text{tot}(E) = \sum_{i=1,2} G_\text{i} \text{Re}\left[\frac{|E|}{\sqrt{E^2 - \Delta_i^2(E)}}\right]\text{,}
\end{equation}
where $G_\text{i}$ is the weighing factor for the two bands $i = \left\{1, 2\right\}$, and $\Delta_i(E)$ is the energy-dependent gap magnitude, which is calculated self-consistently from:
\begin{equation}\label{eq:gap-self-consistent}
    \Delta_i(E) = \frac{\Delta_i^0 + a_{ij}\Delta_j(E)/\sqrt{\Delta_j^2(E) - E^2}}{1 + a_{ij}/\sqrt{\Delta_j^2(E) - E^2}}\text{.}
\end{equation}
In \cref{eq:gap-self-consistent}, $\Delta_i^0$ is the intrinsic gap in band $i$ and $a_{ij}$ quantifies the coupling between the two bands $i$ and $j$ ($i \neq j$).
To include (thermal) broadening effects, \cref{eq:mcmillan} was convoluted by the derivative of the Fermi-Dirac equation with an effective temperature $T_\text{eff}$.
As shown in \cref{fig:sc-gap-fits}, we obtain good agreement with the experimental data.
We find an effective temperature of ca. \qty{500}{mK} for all datasets.
From the parameters of all the fits, which are shown in \cref{tab:sc-gap-fits}, we do not observe a trend that suggests a filling of the SC gap on \crbr{} positions.

\begin{table}[p]
    \centering
    \small
    \begin{tabular}{|l|l|c|c|c|c|c|c|c|}
    \hline
    \rowcolor{lightgray} Panel & Position & $T_\text{eff}$ (K) & $\Delta_1^0$ (meV) & $G_1$ (nS) & $a_{12}$ (meV) & $\Delta_2^0$ (meV) & $G_2$ (nS) & $a_{21}$ (meV) \\ \hline\hline
    \rowcolor[HTML]{e0e0e0} (a) & \nbse{} & 0.49 & 1.32 & 26 & 0.54 & 0.0 & 4.5 & 3.37 \\
    \rowcolor[HTML]{e0e0e0} (b) & \crbr{} & 0.50 & 1.31 & 26 & 0.53 & 0.0 & 4.2 & 3.55 \\ \hline  
    \rowcolor[HTML]{e8Daff} (c) & \nbse{} & 0.50 & 1.26 & 23 & 0.28 & 0.0 & 9.4 & 3.32 \\
    \rowcolor[HTML]{e8Daff} (d) & \crbr{} & 0.50 & 1.29 & 30 & 0.39 & 0.0 & 4.2 & 3.36 \\ \hline  
    \rowcolor[HTML]{bae6ff} (e) & \nbse{} & 0.56 & 1.26 & 75 & 0.17 & 0.0 & 32.7 & 2.20 \\
    \rowcolor[HTML]{bae6ff} (f) & \crbr{} & 0.53 & 1.24 & 104 & 0.27 & 0.1 & 7.7 & 2.03 \\ \hline  
    \rowcolor[HTML]{a7F0ba} (g) & \nbse{} & 0.52 & 1.23 & 32 & 0.35 & 0.0 & 0.0 & 1.96 \\
    \rowcolor[HTML]{a7F0ba} (h) & \crbr{} & 0.53 & 1.22 & 28 & 0.27 & 0.0 & 0.3 & 1.97 \\ \hline  
    \end{tabular}
    \caption{
        \label{tab:sc-gap-fits}
        Optimized parameters obtained by fitting the McMillan model to \didv{} spectra of the SC gap on \nbse{} and \crbr{} positions.
        The first column (`Panel') refers to the panels in \cref{fig:sc-gap-fits}, where the data and the fits are shown.
        The parameters are introduced in \cref{eq:mcmillan,eq:gap-self-consistent}.
    }
\end{table}

\end{appendix}
\clearpage
\bibliography{CrBr3-NbSe2}

\end{document}